\documentstyle[revtex]{aps}
\begin{document}
\begin{center}
\Large\bf
Investigation  of  an  angular distribution of protons in
peripheral and central nucleus-nucleus collisions at the momentum of
4.2 A GeV/c.

\vspace{0.5cm}
\large
M.K. Suleymanov$^{*)}$,O.B. Abdinov, Z.Ya. Sadigov

\large\it
Institute of Physics, Academy of Sciences of Azerbaijan
Republic, Baku

$^{*)}$ E-mail: mais@sunhe.jinr.ru

\vspace{0.5cm}

\large
N.S. Angelov,A.S. Vodopyanov, A.A. Kuznetsov

\large\it
Joint Institute for Nuclear Research, Dubna, Moscow region,Russia

\end{center}
\large
\vspace{0.5cm}

{\bf Abstract}

The experimental results on the relation between the number of events, the
angular distributions of protons and full number of protons are presented
for ${}^{12}CC$-interactions at the momentum of 4.2 A GeV/c. The influence of
nuclear fragmentation process on the results is also considered. The obtained results confirm the assumption that there exist the critical phenomena
among  the central collisions and it is necessary to use a percolation approach for the full description of the central collisions.

\vspace{0.5cm}

{\bf 1. Introduction}

At present there are many papers in which the processes of  nuclear
fragmenatation~[\cite{j1}] and the processes of a total disintegration of
nuclei~[\cite{j2}] ( or central collisions) are considered as the critical
phenomena and  for their description a percolation approach was
proposed.

Some experimental results  obtained in the region of high
energy~[\cite{j4}-\cite{jj}] clearly demonstrate the
existence of the regime change points in the behaviour of  different
characteristics of secondary particles and the events depending on centrality
degree of collisions(in different papers the number of all protons~[\cite{j5}], the number of multicharged fragments~[\cite{j4}] and other parameters were considered for the latter).

We believe that if the observed regime change points are connected with
the appearance and the decay of percolation claster then with the increase of
the centrality degree  of collisions the behaviour of the
secondary particles and the events characteristics depending on centrality
degree of collisions must also depend on the number of  fragments in the
event, as the percolation claster could be a source of nuclear fragments.

The purpose of our investigation is to test this idea experimentally.

\vspace{0.5cm}

{\bf 2. Experiment}

The experimental data have been obtained from the 2-m propane bubble
chamber of LHE, JINR. We used 20407 ${}^{12}CC$- interactions at the
momentum of 4.2 A GeV/c ( for methodical details see~[\cite{j6}]) .

To reach the purpose  we  investigated a number of the events
depending on the variable $Q$.  To determine the values of  $Q$ two
variants were considered.  In the first variant the values  of
$Q$ were determined as $Q=n_{\pi^+}+N_p-n_{\pi^-}$. Here
$n_{\pi^+}$,$n_{\pi^-}$ and $N_p$ are the number of identified
$\pi^+$ -, $\pi^-$ - mesons  and protons respectively (in figures these
points are denoted as the empty starlets).In that determination $Q$ is a number
of all the protons in an event without taking into account a remainder
of nuclei.  In the second variant the values of $Q$ were determined as
$Q=N_{+}-n_{\pi^-}$. Here $N_{+}$ are  charges  of  all the positively
charged particales in an event including  nuclear fragments
(in figures these points are denoted as the full starlets).
In that determination Q is a summary charge of an event.

\vspace{0.5cm}

{\bf 3. Experimental results.}

{\bf 3.1. Q-dependence of the events number.}

The distributions of the events depending on $Q$ are shown in fig.1a,b.
It is seen that with the enclusion of fragments number to determine $Q$ the form
of distributions sharply changes and  has a two--steps
structure(fig.1a, full starlets).

In fig.1b are shown the Q-dependences of the events  for the calculation
data obtained from the quark-gluon string model~[\cite{j7}] (QGSM) without
nuclear fragments. The empty starlets correspond to the cases when the
stripping protons were not taken into account and the full starlets correspond
to the cases when the stripping protons were included.
It is seen that the form of the distribution
strongly differs from the experimetal one in fig.1a.
There is no the two-steps structure in this figure.  Therefore we can
assert that this difference is connected with the existence of
fragments in $^{12}CC$ - interactions.

Thus, the  results demonstrate that the influence of  nuclear
fragmentation process in the behaviour of different
characteristics of the events depending on $Q$ has a critical character.
We suppose this result to be connected with  percolation clusters.

\vspace{0.5cm}

{\bf 3.2. Q-dependence of the protons angular distributions.}

To confirm the existence of percolation claster  we analysed
the angular spectrums of identified protons depending on $Q$ and
the number of fragments. We have obtained
 ${N_i}={N_i'\over \sum_{i=1}^{J} N_i'}={f(\cos\theta_i)}$ (here $N_i'$
are a number of the protons at an emission angle of $\theta_i$
and  $J$ is a total number of the protons in an event) for the following groups of events
 depending on $Q$: $Q\le 5$(this is  peripheral collisions-$N_1$);
$Q=6-7(N_2); 8-9(N_3);\ge 10$(this is  central collisions-$N_4$).

To investigate $Q$-dependences of $N_i$-functions we used the following
quantities: ${f_1}  =  {{N_4-N_3}\over {N_4+N_3 }}~~;
{f_2}  =  {{N_4-N_2}\over { N_4+N_2 }}~~;
{f_3} =  {{N_4-N_1}\over { N_4+N_1 }}$.~~~
To investigate the values of $f_i$ depending on the number of
fragments we also used(as in~(3.1)) two variants to determine the variable $Q$.

Fig.2  shows the $Q$-dependences of $f_1,f_2$ and $f_3$ as a function of
$\cos \theta_i$. As well as in fig.1 the
empty starlets corrisponds to the cases when the nuclear fragments
were not taken into account and the full starlets -- the cases when the
nuclear fragments were included. It is seen that these
distributions difffer for the  two different variants of
$Q$-determination. We see the increase of $f_3$ with
the increase of $\cos \theta_i$ i.e. an
additional production of protons in this interval. We suppose the
percolation claster to be a source of additional protons in this
interval.

\vspace{0.5cm}

{\bf 4. Summary}

For $^{12}CC$-interaction  the behaviour of  the number of events,
depending on $Q$ also depends on the number of
fragments and has a two--steps form. This form is not
reproduced by the calculated data in the framework of the QGS model which
does not take into account nuclear fragments.
This result as well as the results obtained from the
analisys  of angular distributions of protons in peripheral and
central collisions could be a confirmation of the existence of
percolation clasters.

Finally  we want to say that at GSI, AGS and Nuclotron  energies
this result can signal about the existence of the transition of nuclear
matter from nucleon states to its  mixed ones.
At  RHIC or LHC energies , a similar result could help to detect
"critical" signals of phasetransition  nuclear matter.

\vspace{0.5cm}

{\bf Figure captions}

Fig.1. Q-dependence of the number of events.

Fig.2. Q-dependence of the angular distribution of protons.

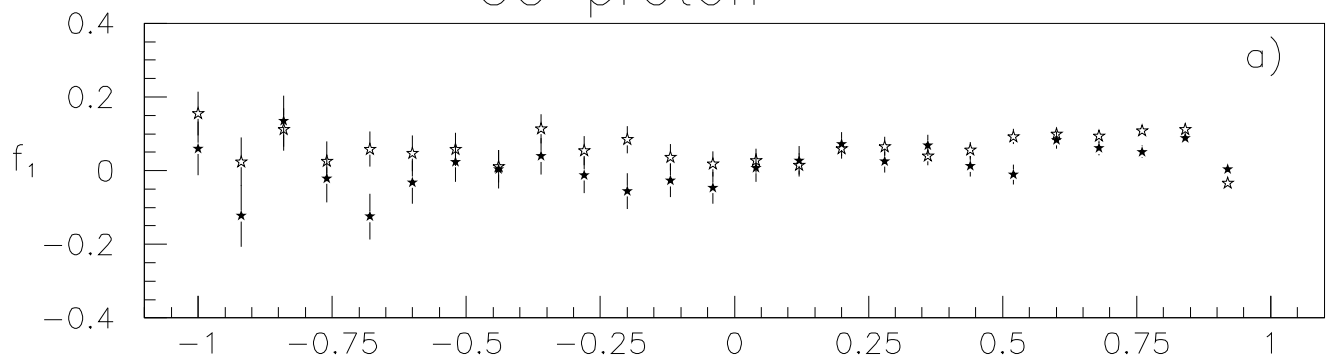
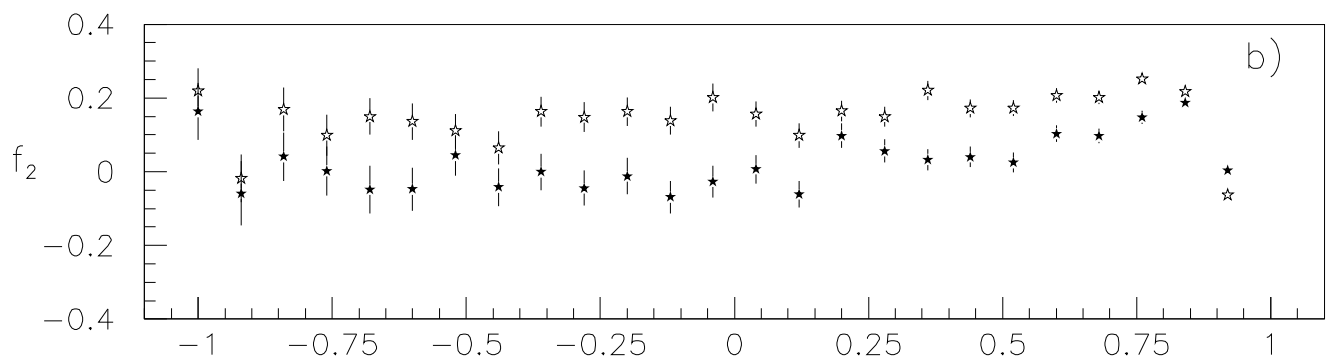
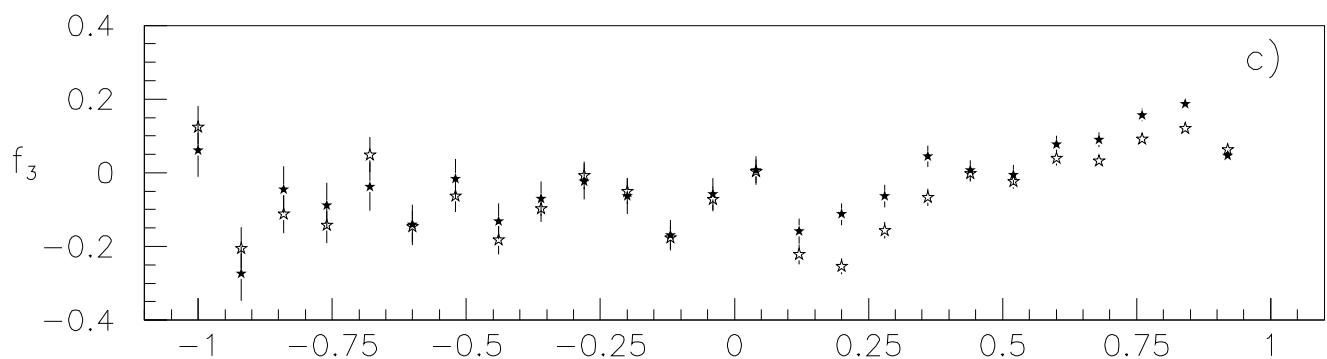
\end{document}